\documentclass[prd,showpacs,floatfix,nofootinbib]{revtex4}
\usepackage{epsfig}

\newcommand{\eq}[1]{Eq.~(\ref{#1})}

\newcommand{\be}{\begin{equation}}
\newcommand{\ee}{\end{equation}}
\newcommand{\bea}{\begin{eqnarray}}
\newcommand{\eea}{\end{eqnarray}}
\newcommand{\ben}{\begin{eqnarray*}}
\newcommand{\een}{\end{eqnarray*}}

\newcommand{\DS}{Dyson-Schwinger }
\newcommand{\BS}{Bethe-Salpeter }
\newcommand{\ST}{Slavnov-Taylor }
\newcommand{\YM}{Yang-Mills }

\newcommand{\w}{\omega}
\newcommand{\e}{\varepsilon}
\newcommand{\al}{\alpha}

\newcommand{\ga}{\gamma}
\newcommand{\G}{\Gamma}
\newcommand{\de}{\delta}

\newcommand{\si}{\sigma}
\newcommand{\Si}{\Sigma}

\newcommand{\la}{\lambda}
\newcommand{\La}{\Lambda}
\newcommand{\ka}{\kappa}

\renewcommand{\div}{\vec{\nabla}}

\newcommand{\s}[2]{{#1}\!\cdot\!{#2}}
\newcommand{\ov}[1]{\overline{#1}}
\newcommand{\dk}[1]{\,\,\,\raisebox{-0.4ex}{\large $\bar{}$}\!\!d\,{#1}\,}

\begin{document}
\title{The Coulomb gauge ghost Dyson-Schwinger equation}
\author{P.~Watson}
\author{H.~Reinhardt}
\affiliation{Institut f\"ur Theoretische Physik, Universit\"at T\"ubingen, 
Auf der Morgenstelle 14, D-72076 T\"ubingen, Deutschland}
\begin{abstract}
A numerical study of the ghost Dyson-Schwinger equation in Coulomb gauge is 
performed and solutions for the ghost propagator found.  As input, lattice 
results for the spatial gluon propagator are used.  It is shown that in 
order to solve completely, the equation must be supplemented by a 
nonperturbative boundary condition (the value of the inverse ghost 
propagator dressing function at zero momentum) which determines if the 
solution is critical (zero value for the boundary condition) or 
subcritical (finite value).  The various solutions exhibit a 
characteristic behavior where all curves follow the same (critical) 
solution when going from high to low momenta until `forced' to freeze out 
in the infrared to the value of the boundary condition.  The 
renormalization is shown to be largely independent of the boundary 
condition.  The boundary condition and the pattern of the solutions can be 
interpreted in terms of the Gribov gauge-fixing ambiguity.  The connection 
to the temporal gluon propagator and the infrared slavery picture of 
confinement is explored.
\end{abstract}
\pacs{12.38.Aw,11.15.Tk}
\maketitle
\section{Introduction}
\setcounter{equation}{0}
The ghost sector in nonperturbative studies of quantum chromodynamics (QCD) 
has invoked considerable interest over the last decade.  One tool for 
studying nonperturbative QCD is the set of \DS equations which, since one 
works in the continuum, is especially suited for discussing the infrared 
behavior of the Green's functions where dynamical singularities may be 
present.  Originally, in Landau gauge it was believed that the ghosts are 
unimportant to the Yang-Mills sector -- the conjecture was that it is the 
three-gluon vertex that is responsible for confinement 
\cite{Mandelstam:1979xd,BarGadda:1979cz}.  However, the work of 
Ref.~\cite{von Smekal:1997vx} turned this conjecture on its head: it was 
shown that the ghost sector dominates the set of \DS equations in the 
infrared, with the three-gluon vertex contributions being subleading.

In order to obtain numerical results, the authors of 
Ref.~\cite{von Smekal:1997vx} introduced infrared fit functions for the 
propagator dressing functions that were characterized by powerlaws (in 
other words, $\sim x^\ka$ for some exponent $\ka$).  Drawing on this idea, 
it was subsequently shown that one can completely characterize the infrared 
behavior of Yang-Mills theory in Landau gauge by studying only the ghost 
sector contributions, with a tree-level ghost-gluon vertex and pure 
powerlaws \cite{Atkinson:1997tu,Atkinson:1998zc}.  The powerlaw theme was 
expanded \cite{Watson:2001yv,Lerche:2002ep,Alkofer:2004it} and it was later 
shown that the tree-level ghost-gluon vertex truncation was reliable 
\cite{Schleifenbaum:2004id}.  Two contemporary reviews can be found in 
Refs.~\cite{Alkofer:2000wg,Fischer:2006ub}.  These results gained some 
degree of popularity not simply because of their simplicity, but also 
because they were in agreement with various descriptions of the confinement 
problem: the results imply positivity violation, such that the propagators 
cannot correspond to physical states and formalized by the Oehme-Zimmermann 
superconvergence relations \cite{Oehme:1979ai}, the results were also in 
agreement with the Kugo-Ojima \cite{Kugo:1979gm} and Gribov-Zwanziger 
\cite{Gribov:1977wm,Zwanziger:1995cv,Zwanziger:1998ez} confinement 
scenarios (at least insofar as they were understood at that time).  The 
lattice results of the day were in good agreement with the powerlaw 
description of the infrared (for a detailed discussion, see for example 
Ref.~\cite{Fischer:2006ub}).  However subsequent lattice studies going 
further into the infrared cast doubt on and eventually seemed to rule out 
altogether \cite{Cucchieri:2007rg,Cucchieri:2008fc} the existence of 
infrared powerlaw solutions for the propagator dressing functions in Landau 
gauge and indicated that the functions should be a finite constant at zero 
momentum (for a detailed discussion, see Ref.~\cite{Cucchieri:2010xr}).  In 
addition, it was pointed out that non-powerlaw solutions also exist to the 
\DS equations (see, for example, Refs.~\cite{Boucaud:2008ji,Boucaud:2008ky} 
and discussions therein).  The coexistence of both types of solution to the 
\DS equations in Landau gauge was studied in Ref.~\cite{Fischer:2008uz}.  
The issue at stake is how the confinement mechanism manifests itself at the 
level of the propagators in Landau gauge and the investigation is ongoing.

The situation in Coulomb gauge studies of nonperturbative QCD is in some 
respects similar.  Coulomb gauge studies come in various guises and indeed, 
the definition of Coulomb gauge is somewhat different depending on which 
formalism one uses.  Currently, the most widely used continuum formalism is 
the canonical formalism, based on the Hamilton density operator \cite{Szczepaniak:2001rg,Feuchter:2004mk,Schleifenbaum:2006bq,Epple:2006hv,
Epple:2007ut,Reinhardt:2008ek,Reinhardt:2009ks}.  In the canonical 
approach, one starts by setting Weyl gauge and then `rotates' into Coulomb 
gauge \cite{Christ:1980ku}, resolving Gauss' law to ensure gauge invariance 
and subsequently minimizing the energy density with an Ansatz for the 
wavefunctional to arrive at a set of Dyson-Schwinger-like equations.  In 
the canonical formalism, there are no temporal degrees of freedom and the 
equations refer only to the equaltime correlation functions.  The 
Dyson-Schwinger-like equations closely resemble the Landau gauge \DS 
equations in a space with one less dimension.  In the canonical approach, 
there is known to be two types of solution for the propagator dressing 
functions \cite{Epple:2007ut} (just like in Landau gauge): critical and 
subcritical and in the critical case, two different values for the powerlaw 
exponent have been found.  The favored solution is the most singular 
solution where the ghost propagator dressing function diverges as 
$1/|\vec{k}|$ in the infrared ($|\vec{k}|\rightarrow0$) because this 
solution has a natural interpretation with regards to physical confinement 
(see the discussions of Refs.~\cite{Reinhardt:2008ek,Reinhardt:2009ks}).  
The corresponding expression for the dressing function of the equaltime 
spatial gluon propagator is very close to Gribov's famous formula 
\cite{Gribov:1977wm} (see later in the text for the explicit form).

There also exists lattice results for Coulomb gauge \cite{Cucchieri:2000gu,
Langfeld:2004qs,Nakagawa:2009zf,Burgio:2008jr,Quandt:2008zj}.  Again, the 
definition of Coulomb gauge plays a role.  In 
Refs.~\cite{Cucchieri:2000gu,Langfeld:2004qs,Nakagawa:2009zf}, the gauge is 
implemented separately on each time-slice and the equaltime correlation 
functions are considered (as in the canonical formalism).  However, these 
results appear to be plagued by scaling violations.  Such scaling 
violations highlight one of the difficulties inherent to Coulomb gauge, 
namely that as yet there is no complete proof of renormalizability 
(although there has been some progress in this direction 
\cite{Zwanziger:1998ez,Baulieu:1998kx,Niegawa:2006hg}).  In an attempt to 
avoid the scaling violations, the authors of 
Refs.~\cite{Burgio:2008jr,Quandt:2008zj} implemented a lattice version of 
Coulomb gauge that includes the temporal degrees of freedom, thereby 
studying the full correlation functions and not their equaltime 
counterparts.  All the lattice studies so far agree on certain results: i) 
the equaltime spatial gluon propagator is well-reproduced by the Gribov 
formula, ii) the ghost propagator is infrared enhanced with the dressing 
function diverging like $1/[\vec{k}^2]^\ka$ and where $\ka\approx0.2-0.25$ 
(at least for as far as these lattice studies go into the infrared regime), 
iii) the temporal gluon propagator is strongly infrared enhanced, probably 
diverging like $1/\vec{k}^4$.  We should however remind the reader of the 
earlier situation for Landau gauge studies (discussed in 
Ref.~\cite{Cucchieri:2010xr}) -- there was a generally agreed infrared 
result for modest lattice sizes (sizes comparable with the current 
Coulomb gauge lattice studies) which was later overturned by much 
larger lattices that probed further into the infrared region.  Therefore, 
one should perhaps take some caution in interpreting the lattice results 
in isolation.  However, one point of note is that the Gribov formula 
for the equaltime spatial gluon propagator in Coulomb gauge vanishes in the 
infrared and this was shown to be true in general on the lattice 
\cite{Zwanziger:1991gz} 
(though there are certain caveats with this, see for example 
Ref.~\cite{Cucchieri:2000gu} for a discussion).  
Taken alongside the results of 
Ref.~\cite{Zwanziger:1991gz}, of the continuum canonical approach 
\cite{Feuchter:2004mk,Epple:2006hv} 
(discussed previously) and in the absence of any direct evidence to the 
contrary it seems fair to say that there is at least a current agreement 
about the equaltime spatial gluon propagator having the Gribov form 
(whether this situation will persist may well prove interesting).  
Certainly, it appears to us reasonable to use the Gribov formula as 
input into the work presented here.

A third approach to Coulomb gauge (and that used in this study) is the 
continuum functional formalism which is based on the Lagrange density of 
QCD.  The importance of the functional formalism in Coulomb gauge lies in 
the recognition that one can reduce the system to physical degrees of 
freedom \cite{Zwanziger:1998ez}.  Further, one can formally show the 
existence of a conserved and vanishing total color charge (along with the 
absence of the infamous Coulomb gauge energy divergences) 
\cite{Reinhardt:2008pr}.  This is crucial because a system that spuriously 
`leaks' color charge cannot be confining.  Because of the inherent 
noncovariance of Coulomb gauge and the fact that one explicitly retains the 
temporal degrees of freedom in the functional formalism, detailed technical 
results are difficult to derive.  However, steady progress is being made: 
the \DS equations have been explicitly derived 
\cite{Watson:2006yq,Watson:2007vc,Popovici:2008ty}, along with the \ST 
identities \cite{Watson:2007vc,Watson:2008fb,Popovici:2010mb} and one-loop 
perturbative results are available 
\cite{Watson:2007mz,Watson:2007vc,Popovici:2008ty}.  These results include 
a study of heavy quarks and the corresponding \BS equation 
\cite{Popovici:2010mb} that motivates an extremely simple quark confinement 
scenario that corresponds to the infrared slavery picture whereby if the 
temporal gluon propagator diverges like $1/\vec{k}^4$ in the infrared (as 
indicated from lattice results, discussed previously) then one has only 
colorless finite energy bound states of quarks and antiquarks with a 
linearly rising potential between them.

In this study, we shall investigate numerical solutions to the Coulomb 
gauge ghost \DS equation within the (second order) functional formalism and 
focus on the issue of critical versus subcritical infrared behavior.  As 
primary input, we utilize the general consensus about the equaltime spatial 
gluon propagator having the Gribov form, as discussed previously.  The 
paper is organized as follows.  In Sec.~\ref{sec:two}, the ghost \DS 
equation and its components will be introduced.  In Sec.~\ref{sec:three}, 
the reduction of the equation to a form suitable for numerical analysis 
will be made explicit, along with a discussion of the asymptotic behavior.  
The numerical results will be presented in Sec.~\ref{sec:four}; the 
comparison of the solutions with available lattice results will also be 
described.  In Sec.~\ref{sec:five}, the results will be discussed and 
arguments put forward about their connection to the Gribov gauge-fixing 
ambiguity.  In Sec.\ref{sec:six}, it will be motivated how the critical 
ghost solution gives rise to a straightforward interpretation of 
confinement.  The paper closes with a summary and outlook in 
Sec.~\ref{sec:seven}.
\section{\label{sec:two}The ghost equation}
\setcounter{equation}{0}

Let us begin by reviewing some basic results for the Green's functions in 
the standard second order functional formalism as applied to Coulomb gauge 
\YM theory.  Throughout this work, we shall use the notations and 
conventions established in 
\cite{Watson:2006yq,Watson:2008fb,Watson:2007vc}.  We work in Minkowski 
space with metric $g_{\mu\nu}=\mbox{diag}(1,-\vec{1})$ (until such time as 
it is necessary to analytically continue to Euclidean space).  Roman 
subscripts ($i,j,\ldots$) denote spatial indices (all minus signs 
associated with covariant/contravariant vectors are explicitly extracted) 
and superscripts ($a,b,\ldots$) denote color indices in the adjoint 
representation (with $N_c$ colors).

\begin{figure}[t]
\vspace{0.5cm}
\includegraphics[width=0.5\linewidth]{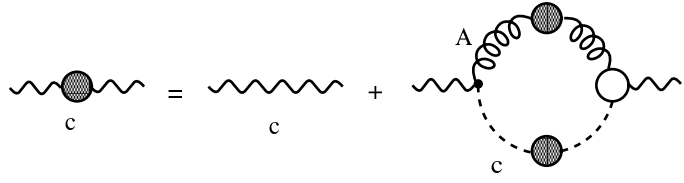}
\caption{\label{fig:ghost0}The unrenormalized \DS equation for the ghost 
two-point proper function, $\G_{\ov{c}c}$.  Filled blobs denote dressed 
two-point functions and empty circles denote dressed proper vertex 
functions.  Wavy lines denote proper functions, springs denote connected 
(propagator) functions and dashed lines denote the ghost propagator.}
\end{figure}

The unrenormalized ghost \DS equation is given by \cite{Watson:2007vc} 
(see also Fig.~\ref{fig:ghost0})
\be
\G_{\ov{c}c}^{af}(k)=
\imath\de^{af}\vec{k}^2+\int\dk{\w}\G_{\ov{c}cAi}^{(0)abc}(k,\w-k,-\w)
W_{\ov{c}c}^{bd}(k-\w)W_{AAij}^{ce}(\w)\G_{\ov{c}cAj}^{dfe}(k-\w,-k,\w),
\label{eq:dsegh0}
\ee
where $\dk{\w}\equiv d^{d+1}\w/(2\pi)^{d+1}$ and with spatial dimension 
$d$.  The ghost propagator ($W_{\ov{c}c}$) and the ghost proper two-point 
function ($\G_{\ov{c}c}$) are decomposed as follows:
\be
W_{\ov{c}c}^{ab}(k)=-\de^{ab}\frac{\imath}{\vec{k}^2}D_{\ov{c}c}(\vec{k}^2),
\;\;\;\;
\G_{\ov{c}c}^{ab}(k)=\de^{ab}\imath\vec{k}^2\G_{\ov{c}c}(\vec{k}^2),
\ee
and the respective (dimensionless) dressing functions obey the usual 
relationship:
\be
D_{\ov{c}c}(\vec{k}^2)\G_{\ov{c}c}(\vec{k}^2)=1.
\ee
Notice that in general, Coulomb gauge propagators and vertices are not 
dependent on the (Lorentz invariant) four-momentum squared 
($k^2=k_0^2-\vec{k}^2$), but rather on the energy squared ($k_0^2$) and the 
momentum squared ($\vec{k}^2$) separately because of the inherent 
noncovariance of Coulomb gauge.  However, we do know that the ghost 
propagator dressing function, $D_{\ov{c}c}$, is strictly independent of the 
energy as a nonperturbative result of the \ST identities 
\cite{Watson:2007vc,Watson:2008fb}.  In addition to the ghost propagator, 
we have the spatial component of the gluon propagator
\be
W_{AAij}^{ab}(k)=
\de^{ab}\frac{\imath}{k^2}t_{ij}(\vec{k})D_{AA}(k_0^2,\vec{k}^2),
\ee
where $t_{ij}(\vec{k})=\de_{ij}-k_ik_j/\vec{k}^2$ is the transverse spatial 
projector.  The above dressing functions ($D_{\ov{c}c}$, $D_{AA}$) reduce 
to unity at tree-level and the explicit one-loop perturbative expressions 
are known \cite{Watson:2007vc,Watson:2007mz}.  The third component of the 
\DS equation, \eq{eq:dsegh0}, is the spatial ghost-gluon vertex which is 
decomposed as follows
\be
\G_{\ov{c}cAi}^{abc}(p_1,p_2,p_3)=
-\imath gf^{abc}\G_{\ov{c}cAi}(p_1,p_2,p_3),
\ee
and at tree-level
\be
\G_{\ov{c}cAi}^{(0)}(p_1,p_2,p_3)=p_{1i}.
\ee
The spatial ghost-gluon vertex has the property that it reduces to the bare 
vertex in the limit of vanishing `in-ghost' spatial momentum 
($|\vec{p}_2|\rightarrow0$) \cite{Watson:2006yq} (as in Landau gauge 
\cite{Marciano:1977su}).  With these decompositions, the ghost \DS 
equation, \eq{eq:dsegh0}, can be rewritten in terms of the dressing 
functions (still unrenormalized and in Minkowski space):
\be
\G_{\ov{c}c}(\vec{k}^2)=
1-\imath g^2N_c\int\frac{\dk{\w}D_{AA}(\w_0^2,\vec{\w}^2)
D_{\ov{c}c}((\vec{k}-\vec{\w})^2)}{\vec{k}^2\w^2(\vec{k}-\vec{\w})^2}
k_it_{ij}(\vec{\w})\G_{\ov{c}cAj}(k-\w,-k,\w).
\label{eq:dsegh1a}
\ee
We immediately notice that in order that the loop integral be independent 
of the external energy ($k_0$), the spatial ghost-gluon vertex (or more 
properly, its contraction within the integral) can only be dependent on the 
energy of the gluon leg ($\w_0$), i.e.,
\be
\G_{\ov{c}cAj}(k-\w,-k,\w)\rightarrow
\G_{\ov{c}cAj}(\vec{k}-\vec{\w},-\vec{k};\w).
\ee
This is entirely consistent with the vertex \ST identities in Coulomb gauge 
\cite{Watson:2008fb}.  Further, we notice that the infrared limit of the 
\DS equation coincides with the known infrared limit of the dressed 
vertex.  This allows us to specify the truncation scheme whereby the fully 
dressed vertex is replaced with its tree-level counterpart:
\be
\G_{\ov{c}cAj}(\vec{k}-\vec{\w},-\vec{k};\w)\rightarrow(k-\w)_j.
\label{eq:gvert0}
\ee
This truncation is known to be robust in Landau gauge studies 
\cite{Watson:2001yv,Lerche:2002ep}, since it reproduces both the infrared 
and ultraviolet behavior of the vertex; the dynamical corrections to the 
dressed vertex in the mid-momentum region have been demonstrated to be very 
modest \cite{Schleifenbaum:2004id}.  Because the ghost couples in both 
Landau and Coulomb gauge to transverse gluon degrees of freedom, we can 
reasonably assume (though this should of course be verified at some stage) 
that this truncation will work well in Coulomb gauge.  Indeed, it has been 
successfully employed in the analyses performed in the canonical approach 
\cite{Feuchter:2004mk,Epple:2006hv,Epple:2007ut}.

Before continuing, let us briefly introduce some aspects of the 
renormalization that are relevant specifically for the ghost \DS equation.  
We assume that the functional approach to Coulomb gauge is 
(nonperturbatively) multiplicatively renormalizable in this study, although 
as mentioned in the introduction, there is currently no complete proof.  
When renormalizing the theory, the process of regularization introduces a 
nontrivial scale.  Recalling the noncovariant nature of Coulomb gauge, we 
initially choose to renormalize at the purely spatial momentum point 
$k_0^2=0$, $\vec{k}^2=\mu$ (where $\mu$ is finite) and assign the fields 
$\vec{A}$, $\ov{c}$, $c$ (we will discuss the temporal component of the 
gluon field, $\si\equiv A_0$, later) and coupling ($g$) the renormalization 
coefficients $\sqrt{Z_A}$, $\sqrt{Z_c}$, and $Z_g$ (the ghost and antighost 
fields share the same coefficient).  We define the renormalized coupling 
via the spatial ghost-gluon vertex; that means that in the effective action 
term for this vertex $\sim g\ov{c}c\vec{A}$ with the corresponding 
renormalization coefficient $\tilde{Z}_A$, we identify 
$\tilde{Z}_A=Z_gZ_c\sqrt{Z_A}$, or
\be
Z_g=\frac{\tilde{Z}_A}{Z_c\sqrt{Z_A}},\;\;\;\;g=\ov{g}Z_g(\mu;\ov{g}),
\ee
where $\ov{g}$ is the renormalized coupling (and is a function of $\mu$ in 
the sense that with a different renormalization point, one would extract a 
different value for the coupling) and the renormalization coefficients are 
all functions of $\mu$ and $\ov{g}$.  We leave the question of the 
regularization procedure (and possible scale) to one side for now.  For the 
propagator dressing functions we then have
\be
D_{AA}(k_0^2,\vec{k}^2;g)=
Z_A(\mu;\ov{g})D_{AA}^R(k_0^2,\vec{k}^2;\mu,\ov{g}),
\;\;\;\;
D_{\ov{c}c}=Z_cD_{\ov{c}c}^R
\ee
(we shall leave aside the common arguments for notational convenience where 
appropriate).  Conventionally, all renormalized propagator dressing 
functions are defined via the renormalization point: 
$D^R(k_0^2=0,\vec{k}^2=\mu;\mu,\ov{g})=1$ (although this will not be 
explicitly required in this work).  As mentioned above, we define the 
renormalized coupling via the spatial ghost-gluon vertex.  To be more 
precise, we demand that at the renormalization point 
($p_3^0=0$, $\vec{p}_2=0$, $\vec{p}_1^2=\vec{p}_3^2=\mu$) the renormalized 
vertex is bare, i.e.,
\be
\G_{\ov{c}cAi}^R(p_3^0=0,\vec{p}_2=0,\vec{p}_1^2=\vec{p}_3^2=\mu;\mu,\ov{g})
=p_{1i}.
\ee
However, the unrenormalized vertex is also bare at this point and we have 
that
\be
\tilde{Z}_A(\mu,\ov{g})=1\;\;\Rightarrow Z_g\sqrt{Z_A}Z_c=1.
\label{eq:rginv1}
\ee
Clearly, this is the same situation as in Landau gauge 
\cite{Marciano:1977su}.

As mentioned in the introduction, the spatial gluon propagator in Coulomb 
gauge takes definite meaning only when one specifies within which formalism 
one is working.  In the second order functional formalism used here, 
because there is no gauge restriction on the temporal component of the 
gluon field, the spatial gluon propagator retains its full temporal 
dependence.  The Fourier transform associated with the complete spatial 
gluon propagator reads ($\dk{\vec{k}}\equiv d^d\vec{k}/(2\pi)^d$):
\bea
W_{AAij}^{ab}(x_1^0,\vec{x}_1;x_2^0,\vec{x}_2)&=&
\int\dk{k}
e^{-\imath k_0(x_1^0-x_2^0)+\imath\vec{k}\cdot(\vec{x}_1-\vec{x}_2)}
W_{AAij}^{ab}(k)\nonumber\\
&=&\de^{ab}\int\dk{\vec{k}}
e^{\imath\vec{k}\cdot(\vec{x}_1-\vec{x}_2)}t_{ij}(\vec{k})
\imath\int_{-\infty}^{\infty}\frac{dk_0}{2\pi}e^{-\imath k_0(x_1^0-x_2^0)}
\frac{D_{AA}(k_0^2,\vec{k}^2)}{(k_0^2-\vec{k}^2+\imath0_+)}.
\eea
In the above, we have made use of the translational invariance and inserted 
the Feynman prescription for dealing with the poles of the propagator.  
Note that the multiplicative renormalizability of the theory 
(discussed above) refers to the complete propagator.  Now let us further 
consider the equaltime case $x_1^0=x_2^0$.  This is the spatial gluon 
propagator that one considers within the canonical  (Hamiltonian) formalism 
\cite{Feuchter:2004mk} and on the lattice where Coulomb gauge is fixed 
separately on each time-slice 
\cite{Cucchieri:2000gu,Langfeld:2004qs,Nakagawa:2009zf}.  Thus, we require 
that the integral (which will play a crucial role in the ghost self-energy 
studied in detail in the next Section)
\be
D_{AA}^T(\vec{k}^2)=
\imath\int_{-\infty}^{\infty}\frac{dk_0}{2\pi}
\frac{D_{AA}(k_0^2,\vec{k}^2)}{(k_0^2-\vec{k}^2+\imath0_+)}
=\int_{-\infty}^{\infty}\frac{dk_4}{2\pi}
\frac{D_{AA}(-k_4^2,\vec{k}^2)}{(k_4^2+\vec{k}^2)}
\label{eq:datdef}
\ee
(the second form arising from the Wick rotation: 
$k_0\rightarrow\imath k_4$, which we assume to be possible) be 
well-defined.  This places a weak restriction on the large (Euclidean) 
energy behavior of the dressing function $D_{AA}$ for arbitrary, finite, 
momentum $\vec{k}^2$.  One important observation about this restriction 
concerns the perturbative limit and highlights a difficult technical 
problem associated with the functional approach to Coulomb gauge.  
Perturbation theory is only valid in the vicinity of the renormalization 
point (chosen here to be purely spacelike, i.e., $k_0^2=0$); away from 
this point, the corrections grow logarithmically.  When integrating over 
the energy, one is extending into the high (Euclidean) energy region where 
the perturbative result is invalid.  Thus for example, in order to compare 
a numerical result for $D_{AA}$ with the (large momentum) perturbative 
expression for $D_{AA}^T$ in the canonical approach, one already requires 
nonperturbative information about $D_{AA}$, although the asymptotic series 
expansions still agree at one-loop \cite{Campagnari:2009km}.  Similar 
observations apply for the temporal component of the gluon propagator 
\cite{Cucchieri:2000hv}.

In the next section, it will be seen that in order to solve the ghost 
\DS equation, one requires $D_{AA}^T$ as input, as opposed to $D_{AA}$.  
As was discussed in the introduction, there is a current consensus that 
$D_{AA}^T$ takes a form consistent with the Gribov formula.  For 
definiteness however, let us begin by discussing the results of one 
particular lattice study of $D_{AA}$ \cite{Burgio:2008jr} (to our knowledge 
the only calculation of this quantity to date).  These results can be 
summarized as follows:
\be
D_{AA}(-k_4^2,\vec{k}^2)=
\left[1+\frac{k_4^2}{\vec{k}^2}\right]^{\al(g_l)}
\frac{\vec{k}^2}{\sqrt{\vec{k}^4+m^4}}.
\ee
The lattice data are in Euclidean space, with the gauge group $SU(2)$ and 
with a lattice coupling $g_l$.  The scale 
$m\approx0.88\mbox{GeV}\sim2\sqrt{\si_w}$ where $\si_w$ is the Wilsonian 
string tension.  It is observed that $m$ is not dependent on the 
renormalization scale.  The exponent $\al(g_l)$ has the following behavior:
\be
\al(g_l)\approx\left\{
\begin{array}{ll}0,&g_l\;\mbox{`small'}\\1,&g_l\;\mbox{`large'}\end{array}
\right..
\ee
Importantly, the temporal behavior of the dressing function factorizes into 
a dimensionless function and one obtains with \eq{eq:datdef}
\be
D_{AA}^T(\vec{k}^2)=
\frac{\G(1/2-\al(g_l))}{\G(1/2)\G(1-\al(g_l))}
\frac{1}{2}\frac{\sqrt{\vec{k}^2}}{\sqrt{\vec{k}^4+m^4}}.
\ee
The temporal behavior of $D_{AA}$ collapses to an overall constant 
prefactor which will be seen in the next section to be largely irrelevant.  
The restriction, following the definition of $D_{AA}^T$, \eq{eq:datdef}, 
on the large (Euclidean) energy behavior of $D_{AA}$ is related to whether 
or not this constant prefactor is finite or divergent.  Clearly, the 
`small' coupling result: $\al(g_l)=0$ is more pertinent since we are 
interested in the physical regime where the (renormalized) coupling is 
small and for `large' coupling, the system undergoes various nontrivial 
phase transitions on the lattice.  Let us thus write
\be
D_{AA}^T(\vec{k}^2)=\frac{1}{2}\frac{\sqrt{\vec{k}^2}}{\sqrt{\vec{k}^4+m^4}}
\label{eq:lattin}
\ee
such that $D_{AA}^T\rightarrow1/(2|\vec{k}|)$ as 
$\vec{k}^2\rightarrow\infty$ as an arbitrary normalization condition for 
now (it corresponds to the tree-level result in the canonical formalism 
\cite{Feuchter:2004mk,Campagnari:2009km}).

The expression \eq{eq:lattin} is of course Gribov's original formula for 
the equaltime spatial gluon propagator \cite{Gribov:1977wm} and as 
discussed in the introduction, there is a current consensus that this is 
the correct (or close to correct) result.  Notice that the above lattice 
result is already renormalized and there is no reference to the 
renormalization scale (as mentioned, the Gribov scale $m$ is observed to be 
independent of the renormalization scale) except implicitly through the 
normalization.  Interestingly (and unlike in Landau gauge), for large 
momenta there is no evidence on the lattice for a perturbative gluon 
anomalous dimension within $D_{AA}^T$, i.e., the coefficient of the 
perturbative logarithms is consistent with zero \cite{Burgio:2008jr} and 
this is consistent with the other lattice studies 
\cite{Cucchieri:2000gu,Langfeld:2004qs,Nakagawa:2009zf}.
\section{\label{sec:three}Analytic framework}
\setcounter{equation}{0}

Inserting the truncated form for the dressed ghost-gluon vertex, 
\eq{eq:gvert0}, and rewriting in terms of renormalized two-point dressing 
functions, the ghost \DS equation, \eq{eq:dsegh1a} now reads
\be
\G_{\ov{c}c}^R(\vec{k}^2)=
Z_c-\ov{g}^2N_c\int
\frac{\dk{\vec{\w}}D_{\ov{c}c}^R((\vec{k}-\vec{\w})^2)}
{\vec{k}^2(\vec{k}-\vec{\w})^2}
k_ik_jt_{ij}(\vec{\w})\imath\int_{-\infty}^{\infty}
\frac{d\w_0}{2\pi}
\frac{D_{AA}^R(\w_0^2,\vec{\w}^2)}{\left(\w_0^2-\vec{\w}^2+\imath0_+\right)}.
\ee
Because the truncated form of the vertex is not dependent on the energy, we 
immediately recognize $D_{AA}^T$, given by \eq{eq:datdef} and inserting our 
lattice input, \eq{eq:lattin}, we then have
\be
\G_{\ov{c}c}^R(\vec{k}^2)=
Z_c-\frac{1}{2}\ov{g}^2N_c\int
\frac{\dk{\vec{\w}}\sqrt{\vec{\w}^2}D_{\ov{c}c}^R((\vec{k}-\vec{\w})^2)}
{\vec{k}^2\sqrt{\vec{\w}^4+m^4}(\vec{k}-\vec{\w})^2}k_ik_jt_{ij}(\vec{\w}).
\label{eq:dsegh1}
\ee
Recall that $D_{AA}^T$ was arbitrarily normalized to remove the coefficient 
(the combination of gamma-functions) resulting from the energy dependence 
of $D_{AA}$.  Were we to relax this normalization condition, we would have 
some overall (constant) prefactor for the ghost self-energy.

Before continuing, let us consider the perturbative treatment of 
\eq{eq:dsegh1} at one-loop (obtained by setting $m=0$ and 
$D_{\ov{c}c}$ within the integral to unity).  Using dimensional 
regularization, we get the standard result \cite{Watson:2007vc}
\be
\G_{\ov{c}c}^R(\vec{k}^2)=
Z_c+\frac{4}{3}N_c\frac{\ov{g}^2}{(4\pi)^2}
\ln{\left(\frac{\vec{k}^2}{\mu}\right)}-
\frac{4}{3}N_c\frac{\ov{g}^2}{(4\pi)^2}
\left[\frac{1}{\e}-\ga+\frac{7}{3}-2\ln{2}+\ln{(4\pi)}\right]
\ee
from which one can identify ($\la\equiv\ov{g}^2/(4\pi)^2=\al_s/(4\pi)$)
\be
\G_{\ov{c}c}^R(\vec{k}^2)=
1+\frac{4}{3}\la N_c\ln{\left(\frac{\vec{k}^2}{\mu}\right)}+{\cal O}(g^4).
\label{eq:ghpert}
\ee
At this level in perturbation theory, one need not consider the running of 
the coupling (i.e., $\ov{g}=g$).  Assuming that $\al_s$ is small and 
$\vec{k}^2$ is close to $\mu$ so that the logarithm is also small, the 
leading order expression can be resummed and inverted to give
\be
D_{\ov{c}c}^R(\vec{k}^2;\mu,\ov{g})=
\left(\frac{\vec{k}^2}{\mu}\right)^{\ga_g},
\;\;\;\;
\ga_g=-\frac{4}{3}\la N_c,
\ee
where $\ga_g$ is the leading order expression for the ghost anomalous 
dimension.

As it stands, \eq{eq:dsegh1} contains logarithmically UV-divergent pieces.  
To eliminate the divergences, we use a nonperturbative subtraction.  Taking 
the finite renormalization scale $\mu$ as our subtraction point, we then 
have
\bea
\lefteqn{\G_{\ov{c}c}^R(\vec{k}^2)=
\G_{\ov{c}c}^R(\vec{k}^2=\mu)}\nonumber\\&&
-\frac{1}{2}\ov{g}^2N_c\int
\frac{\dk{\vec{\w}}\sqrt{\vec{\w}^2}}{\sqrt{\vec{\w}^4+m^4}}
\left\{\frac{D_{\ov{c}c}^R((\vec{k}-\vec{\w})^2)}{(\vec{k}-\vec{\w})^2}
\frac{k_ik_j}{\vec{k}^2}t_{ij}(\vec{\w})
-\left.\frac{D_{\ov{c}c}^R((\vec{k}-\vec{\w})^2)}{(\vec{k}-\vec{\w})^2}
\frac{k_ik_j}{\vec{k}^2}t_{ij}(\vec{\w})\right|_{\vec{k}^2=\mu}\right\}.
\eea
One can see that (as long as $D_{\ov{c}c}^R(x)$ is smoothly varying for 
$\vec{k}^2,\mu\ll x\rightarrow\infty$) the integrand then falls faster than 
$1/|\vec{\w}|^3$ at large $|\vec{\w}|$ to ensure convergence.  Translating 
the integration variable $\vec{\w}\rightarrow\vec{k}-\vec{\w}$, the 
integral can be conveniently rewritten using a UV-cutoff ($\La$) and 
introducing some notation, we write:
\bea
x&=&\vec{k}^2,\;\;\;\;y=\vec{\w}^2,\;\;\;\;
\s{\vec{k}}{\vec{\w}}=\sqrt{xy}z,\nonumber\\
\int\dk{\vec{\w}}&\rightarrow&\frac{2}{(4\pi)^2}
\int_{-1}^1dz\int_0^\La dy\sqrt{y},\;\;\;\;\La\rightarrow\infty,
\eea
to give (we reinsert the renormalization scale dependence of the functions 
and drop the sub- and superscript notation for $\G_{\ov{c}c}^R$ for clarity)
\be
\G(x;\mu)=\G(\mu;\mu)-\la N_c\int_0^\La\frac{dy}{y}\G(y;\mu)^{-1}
\left[I(x,y;m)-I(\mu,y;m)\right],
\label{eq:ghdse1}
\ee
where the (angular integral) function $I$ reads
\be
I(x,y;m)=\int_{-1}^{1}dz(1-z^2)
\left[1+\frac{x}{y}-2\sqrt{\frac{x}{y}}z\right]^{-1/2}
\left[(1+\frac{x}{y}-2\sqrt{\frac{x}{y}}z)^2+\frac{m^4}{y^2}\right]^{-1/2}.
\label{eq:angint1}
\ee
Note that after the UV-divergence has been subtracted, the above 
representation of the self-energy integral is exact, as long as 
$\La\rightarrow\infty$.  Also, when the Gribov scale $m=0$, one recovers 
the free gluon propagator.  For $m=0$, the angular integral can be 
performed analytically and
\be
I(x,y;m=0)=
\frac{4}{3}
\left[\Theta(x-y)\left(\frac{y}{x}\right)^{3/2}+\Theta(y-x)\right].
\ee
With this, one recovers the perturbative result, \eq{eq:ghpert}, when 
expanding in $\al_s$:
\be
\G(x;\mu)=\G(\mu;\mu)+\frac{4}{3}\la N_c\ln{\left(\frac{x}{\mu}\right)}.
\ee

At this stage, one can make a preliminary infrared analysis of 
\eq{eq:ghdse1}.  For fixed $\ov{g}^2$ and $m$, given that the dressing 
functions are dimensionless, one can make the initial Ansatz that 
$\G(x;\mu)=\G(x/\mu)$, disregarding for now the influence of the Gribov 
scale $m$.  One can also infer from the sign of the perturbative logarithm 
that $\G(x/\mu)$ increases with $x$, discounting for now the possibility 
that the function has some turning point (this will be seen numerically not 
to be the case).  Thus, we can write for $x/\mu\rightarrow0$
\be
\G(x/\mu)=a_0+a_1\left(\frac{x}{\mu}\right)^{\al},\;\;\;a_0,a_1\geq0,\;\al>0.
\ee
The constant $a_0$ is determined by the constant part of \eq{eq:ghdse1}.  
To get the powerlaw solution, we assume that this constant vanishes 
(meaning that $\G(x/\mu)\sim(x/\mu)^{\al}$) and rewrite \eq{eq:ghdse1} in 
terms of $x/\mu$ and $y/\mu$:
\be
\G(x/\mu)=\G(1)-\la N_c\int_0^\La\frac{dy}{y}\G(y/\mu)^{-1}
\left[I(x/\mu,y/\mu;m/\sqrt{\mu})-I(1,y/\mu;m/\sqrt{\mu})\right],
\ee
noting that the function $I$ is dimensionless.  The integrals 
(angular and radial) will not alter the exponent of the infrared expansion 
since they are independent of $x$ and $\mu$, so to get the leading powerlaw 
relationship for the exponent, one must simply remove $dy/y$ and replace 
$y/\mu$ in the integral with $x/\mu$, similarly removing $dz$ and setting 
$z=0$ (in other words, dimensional analysis!) and then let 
$x/\mu\rightarrow0$.  Explicitly
\be
\G(x/\mu)\sim\G(x/\mu)^{-1}
\left[I(x/\mu,x/\mu;m\sqrt{\mu})-I(1,x/\mu;m\sqrt{\mu})\right]
\ee
where
\be
I(x/\mu,x/\mu;m/\sqrt{\mu})-I(1,x/\mu;m\sqrt{\mu})\sim
\left[\left(\frac{x}{\mu}\right)^{1}-\left(\frac{x}{\mu}\right)^{3/2}\right]
\left[\frac{x^2}{\mu^2}+\frac{m^4}{\mu^2}\right]^{-1/2}
\sim\left(\frac{x}{m^2}\right)
\ee
so that the powerlaw relationship reads
\be
\left(\frac{x}{\mu}\right)^{\al}\sim
\left(\frac{x}{\mu}\right)^{-\al+1}\frac{\mu}{m^2}
\ee
or $\al=1/2$ if one counts the powers of $x$ ($=\vec{k}^2$).  If one counts 
instead the powers of $\mu$, then one has $\al=0$ which is the constant 
solution.  To summarize, we would expect that if the dressing function 
$\G$ vanishes in the infrared, there is a leading powerlaw behavior given 
with the exponent $\al=1/2$.  This is in agreement with the results 
obtained within the canonical formalism 
\cite{Feuchter:2004mk,Epple:2006hv,Epple:2007ut}.  We will see later that 
the numerical result (derived independently of this crude analysis) 
confirms this behavior.

Returning to \eq{eq:ghdse1}, for $m\neq0$ the $\mu$-subtracted equation is 
still not suitable for numerical analysis because for general 
$\G^{-1}(y;\mu)$ within the integral, the integral can become negative - 
this is obvious from the (not resummed) perturbative result, 
\eq{eq:ghpert}, (for which $\G^{-1}(y;\mu)=1$) when $x\ll\mu$, where 
$\ln{(x/\mu)}\rightarrow-\infty$ \footnote{Demanding that such a negative 
integral be explicitly excluded perturbatively led to the original 
derivation of the Gribov factor used here as input \cite{Gribov:1977wm}.}.  
Therefore, an iterative procedure to solve will automatically fail since 
we require $\G^{-1}(y;\mu)$ to be finite and positive within the 
integration domain (it can have at most an integrable singularity at 
$y=0$).  To avoid such a situation, we subtract \eq{eq:ghdse1} once more 
but now at the position $x=0$ to get
\be
\G(x;\mu)=
\G(0;\mu)+\la N_c\int_0^\La\frac{dy}{y}\G(y;\mu)^{-1}
\left[I(0,y;m)-I(x,y;m)\right].
\ee
Note that the above equation is still formally renormalized at the scale 
$\mu\neq0$.  The bracketed combination of angular integrals is however 
positive definite for all $x>0$ (and zero for $x=0$).  The dependence of 
the above equation on the original renormalization point is now only 
implicitly given within the coupling (and the normalization condition for 
$D_{AA}^T$).  Further, one can rescale the variables 
$x\rightarrow xm^2$, $y\rightarrow ym^2$ such that the new (dimensionless) 
$x,y$ are in units of $[\mbox{GeV}]^2/m^2$.  This is a consequence of the 
fact that the form of the renormalized gluon propagator was fixed from the 
beginning.  The rescaling proportional to $m^2$ also means that the 
comparison with the leading order perturbative result (where $m=0$) will at 
most be for asymptotically large $x$.  Thus, for fixed $\al_s=4\pi\la$ as 
input, the solution is independent of $\mu$ and we can write
\be
\G(x)=\G(0)+\la N_c\int_0^\La\frac{dy}{y}\G(y)^{-1}
\left[\frac{4}{3}\frac{y}{\sqrt{1+y^2}}-I(x,y;1)\right],
\label{eq:ghdse2}
\ee
where we have evaluated $I(0,y;1)$.  There are now only two input 
parameters, $\la N_c$ and $\G(0)$.  For definiteness, we choose the 
physical values $\al_s(M_z)=0.1187$ and $N_c=3$ and consider the solution 
with a range of values for $\G(0)$.  From the above equation, it is now 
obvious that the overall normalization of  $D_{AA}^T$ can be absorbed into 
an irrelevant constant prefactor for $\G(x)$.  (This observation also 
applies to the comparison of the case $N_c=2$ with $N_c=3$.)  Thus we 
observe that the only relevant information about the original input gluon 
propagator is that it had the Gribov form at equal times, the values of the 
Gribov scale and normalization condition will not affect the overall 
features of the ghost solution.  The numerical solutions to \eq{eq:ghdse2} 
will be discussed in the next section.

Having discussed the renormalized ghost \DS equation, let us now briefly 
discuss the associated renormalization coefficient $Z_c(\mu,\ov{g})$.  
Inserting the form for the spatial gluon propagator dressing function, 
\eq{eq:dsegh1} can be written in terms of the angular integral and the 
rescaled variables in the following way
\be
Z_c(\La,[\al_s,\G(0)])=\G(x)+\la N_c\int_0^\La\frac{dy}{y\G(y)}I(x,y,1).
\ee
In the above, we have recognized that $Z_c$ is dependent on $\La$, 
$\la=4\pi\al_s$ and $\G(0)$.  The renormalization scale ($\mu$) dependence 
is implicit within $\al_s$, which is here fixed.  If the \DS equation is 
multiplicatively renormalizable, then $Z_c$ should be observed to be 
independent of $x$ and this is numerically verified (the deviations are of 
the same order as the tolerance for the convergence of the iterative 
solution to \eq{eq:ghdse2}).  It is also seen that $Z_c$ is independent of 
the numerical IR cutoff.  Since $Z_c$ is independent of $x$, then we could 
equivalently set $x=0$ and use the simple form
\be
Z_c(\La,[\al_s,\G(0)])=
\G(0)+\frac{4}{3}\la N_c\int_0^\La\frac{dy}{\sqrt{1+y^2}\G(y)}.
\ee
\section{\label{sec:four}Numerical Results}
\setcounter{equation}{0}

\begin{figure}[t]
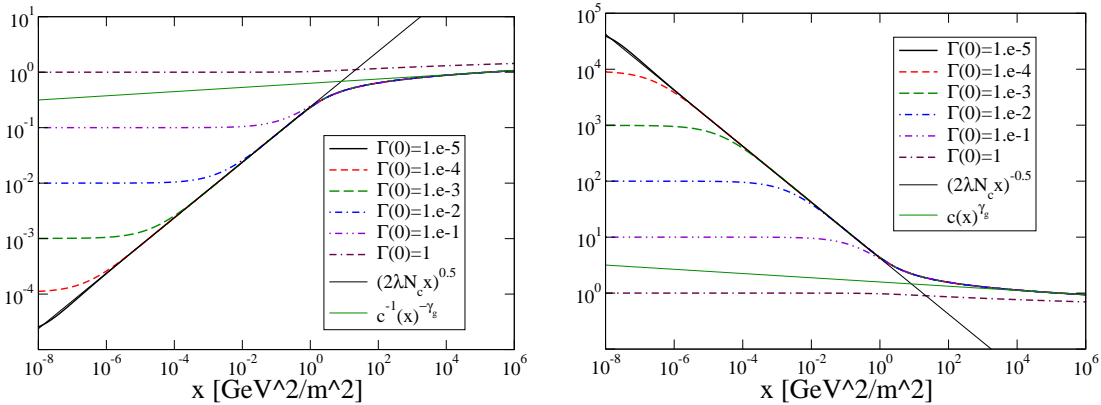

\vspace{0.5cm}
\includegraphics[width=0.39\linewidth]{ghostg00.eps}\hspace{0.5cm}
\includegraphics[width=0.39\linewidth]{ghostd00.eps}
\caption{\label{fig:ghres0}Results for the ghost dressing functions $\G(x)$ 
[left panel] and $D(x)$ [right panel].  See text for details.}
\end{figure}

The results (for $\G(x)$ and $D(x)=\G(x)^{-1}$) are shown in 
Fig.~\ref{fig:ghres0} along with a fit to the supposed infrared powerlaw 
and the asymptotic perturbative result.  Numerically, the solution is stable 
apart for the deep infrared region in the case of small $\G(0)$.  In this 
case one has an integrable singularity and we explicitly do not include the 
deeper level of numerical sophistication required to deal with the problem 
since this would necessarily involve an assumption about the infrared 
behavior of the solution.  However, the features of the solution are 
clearly evident.

In the UV region, all solutions with different $\G(0)<0.1$ tend to the same 
curve (the case $\G(0)=1$ will be discussed presently).  Asymptotically, it 
is seen that these solutions obey
\be
\G(x)^{-1}=D_{\ov{c}c}^R(x)=cx^{\ga_g},
\ee
where $c\approx1.581$ and confirming that one recovers the (resummed) 
perturbative behavior for large $x$.  That all curves have the same 
asymptotic large $x$ behavior suggests that for a large renormalization 
scale $\mu$ (and after scaling variables with $m^2$, $\mu$ would have to be 
asymptotically large), there is a unique value of $\G(\mu)$ (or vanishingly 
small differences between different values) for arbitrary input $\G(0)$.  
We can thus infer that our input $\G(0)$ (at least for moderate values) has 
nothing to do with the perturbative regime and its renormalization -- it is 
genuinely nonperturbative in origin.

\begin{figure}[t]
\vspace{0.5cm}
\includegraphics[width=0.5\linewidth]{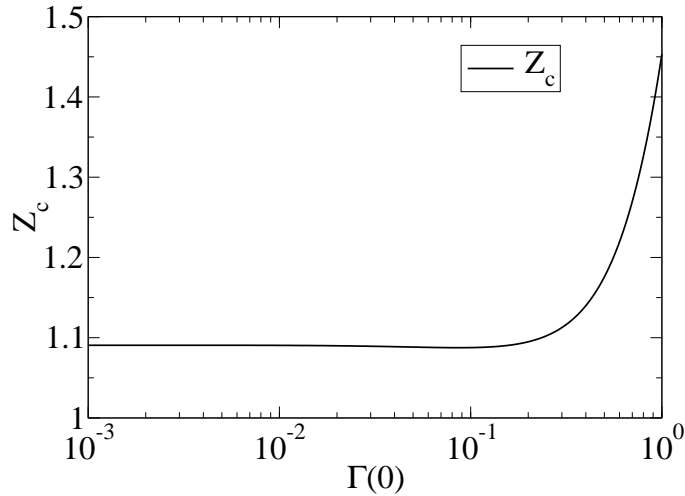}
\caption{\label{fig:zc}Plot of the ghost renormalization coefficient 
$Z_c$ as a function of $\G(0)$.  See text for details.}
\end{figure}

Continuing with the UV region, results for $Z_c$ are given in 
Table~\ref{tab:zc0}.  Effectively, $Z_c\sim1$ and for the $\La$ and $\G(0)$ 
values sampled is slowly varying, increasing with $\La$ like 
$Z_c(\La)\sim\ln{\La}$ as one would have expected from the perturbative 
result (and with fixed spatial gluon input and a tree-level vertex, this is 
not surprising).  Noticing the behavior of $Z_c$ for different values of 
input $\G(0)$, we plot $Z_c$ as a function of $\G(0)$ for fixed 
$\La=10^6[\mbox{GeV}]^2/m^2$ in Fig.~\ref{fig:zc}.  It is seen that $Z_c$ 
is roughly constant for input values less than $\G(0)\sim0.3$ and this 
gives an estimate of the largest value of $\G(0)$ for which the ghost 
solution does not change in the UV.  This maximum value of $\G(0)$ loosely 
corresponds to the situation where the solution curve $\G(x)$ loses contact 
with the common (asymptotic) large $x$ curve in Fig.~\ref{fig:ghres0} (as 
illustrated for the case $\G(0)=1$).  The inference is that there is a 
region of values $0\leq\G(0)<0.3$ for which the input constant $\G(0)$ is 
unimportant to the UV.  We shall discuss this later in more detail.
\begin{table}
\begin{tabular}{|c||c|c||c|c||c|c||c|c|c|c|}\hline
$\G(0)$&$1$    &$1$    &$10^{-2}$ &$10^{-2}$ &$10^{-4}$ &$10^{-4}$ 
&$0$    &$0$    &$0$    &$0$    
\\\hline\rule[-2.4ex]{0ex}{5.5ex}
$\La$  &$10^6$ &$10^8$ &$10^6$    &$10^8$    &$10^6$    &$10^8$    
&$10^6$ &$10^7$ &$10^8$ &$10^9$
\\\hline\hline\rule[-2.4ex]{0ex}{5.5ex}
$Z_c$  &$1.45$ &$1.57$ &$1.09$    &$1.24$    &$1.09$    &$1.24$    
&$1.09$ &$1.17$ &$1.24$ &$1.31$
\\\hline
\end{tabular}
\caption{\label{tab:zc0}Numerical results for the renormalization 
coefficient $Z_c$.  Dimensionful quantities are in units of 
$[\mbox{GeV}]^2/m^2$, see text for details.}
\end{table}

Returning to Fig.~\ref{fig:ghres0}, in the infrared region for small 
$\G(0)$ ($<0.1$) the solution has a region for which the naive powerlaw 
solution holds and as $\G(0)$ is decreased, one can see that this region 
extends further into the infrared.  Moreover, extrapolating to the case 
$\G(0)=0$ by eye, one can see that the pure powerlaw solution does exist.  
Let us emphasize that the powerlaw has not been artificially introduced 
into the numerical solution via sophisticated techniques (and for which 
the numerical stability could be improved to an arbitrary degree).  The 
results are obtained using iteration and a direct numerical integration 
grid: the observed infrared behavior is an independent result.  The 
`fitted' (i.e., fitted by eye) form of the powerlaw solution has one input 
parameter - the coefficient (the exponent was fixed from the earlier 
analysis).  Clearly, the coefficient must have the factor $\sqrt{\la N_c}$ 
but that the factor $\sqrt{2}$ works so well is not explained (and given 
the previous discussions about the normalization of $D_{AA}^T$ is not 
really relevant).    To summarize the IR behavior: the most notable feature 
of the various solutions is that when going from the common UV asymptotic 
solution towards small $x$, for a given input boundary condition $\G(0)$ 
the system has a `preferred' dynamical curve corresponding to the powerlaw 
case $\G(0)=0$ until `forced' to deviate to the constant boundary value 
(if $\G(0)\neq0$).

\begin{figure}[t]
\vspace{0.5cm}
\includegraphics[width=0.5\linewidth]{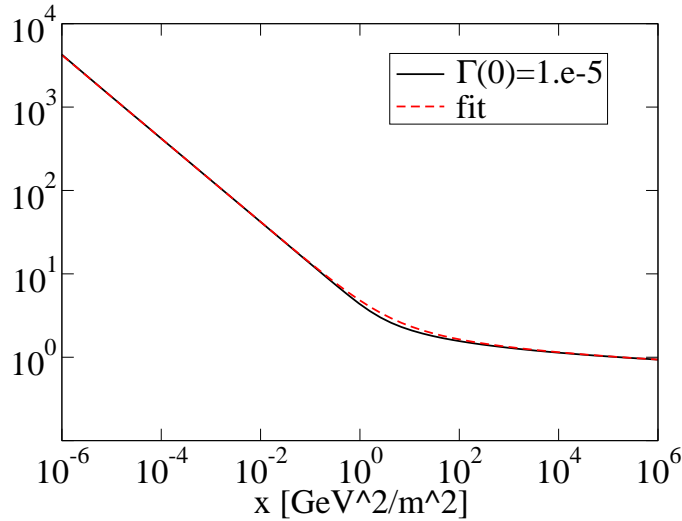}
\caption{\label{fig:gfit}Plot of the ghost propagator dressing function in 
the case $\G(0)=10^{-5}$ (indistinguishable from the powerlaw for the 
scales shown) compared with the fit formula \eq{eq:gfit}.  See text for 
details.}
\end{figure}

In the case of the pure powerlaw solution (numerically here, the lowest 
value of $\G(0)=10^{-5}$), we find that the ghost propagator dressing 
function solution can be well reproduced over the entire momentum range by 
the form
\be
D_{\ov{c}c}(x)=
\frac{1}{\sqrt{2\la N_c}}\sqrt{\frac{(1-a)}{x}+\frac{a}{\ln{(1+x)}}},
\label{eq:gfit}
\ee
with a single parameter, $a\approx0.689$ (we find that $a$ is slightly 
dependent on $\La$, which has here the value $\La=10^6$ in the scaled 
units).  The comparison is shown in Fig.~\ref{fig:gfit}.  One can see that 
the form given in \eq{eq:gfit} will exactly reproduce the known infrared 
behavior because of the small $x$ expansion of the logarithm.  That the 
large momentum behavior is, up to a constant, given by $1/\sqrt{\ln{x}}$ 
was also found in Ref.~\cite{Feuchter:2004mk}.  Recall that it was earlier 
seen that the propagator dressing function behaves as $\sim x^{\ga_g}$ for 
asymptotically large $x$: the $1/\sqrt{\ln{x}}$ behavior represents an 
alternative description to the naive perturbative anomalous dimension and 
indeed appears superior since it is valid over a wide range of momenta.
\subsection{Comparison with lattice results}
We have seen that using a lattice inspired result \cite{Burgio:2008jr} as 
input for the spatial gluon propagator, the ghost propagator dressing 
function either has an infrared divergence characterized by the exponent 
$1/2$, or is constant in the infrared depending on the boundary condition 
$\G(0)$.  This result could be compared to the corresponding lattice 
results.  Various studies 
\cite{Cucchieri:2000gu,Langfeld:2004qs,Nakagawa:2009zf,Quandt:2008zj} have 
reported an infrared divergence, i.e., a propagator dressing function of 
the form $x^{-\ka}$ for small $x$, but with an exponent 
$\ka\approx0.2-0.25$ (all studies agree on the Gribov form for the 
equaltime spatial gluon propagator), apparently in contradiction to the 
results here and from the canonical approach 
\cite{Feuchter:2004mk,Schleifenbaum:2006bq} (where $\ka=1/2$).

The aforementioned lattice studies naturally involve a smallest infrared 
spatial momentum scale $\vec{k}^2/m^2=x_{min}$ and 
$x_{min}\approx0.25-0.5$.  To extract the infrared exponent $\ka$, one 
requires a range of momenta for which the powerlaw relationship holds 
(or the constant $\G(0)$ is clearly visible).  However, taking this range 
to be, for example, one order of magnitude (i.e., $x=[0.25,2.5]$) and 
comparing with the results displayed in Fig.~\ref{fig:ghres0}, one sees 
that all curves (with $0\leq\G(0)<0.3$) are still within the transition 
region where the exponent changes from the perturbative value ($x^{\ga_g}$) 
to the infrared ($x^{-1/2}$).  In other words, it \emph{may} be that the 
lattice results have not yet genuinely probed far enough into the infrared 
to see the full exponent.

\begin{figure}[t]
\vspace{0.5cm}
\includegraphics[width=0.5\linewidth]{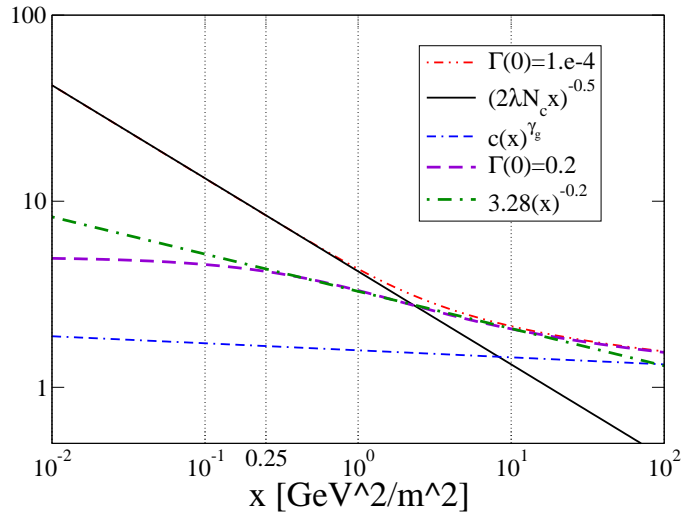}
\caption{\label{fig:latconst}Plot of the ghost propagator dressing function 
in the case $\G(0)=10^{-4}$ (indistinguishable from $\G(0)=0$ on the 
scales shown) and $\G(0)=0.2$ compared with the powerlaws characterized by 
the exponents $\ka=-\ga_g$ (perturbative), $\ka=0.2$ (illustrative lattice 
result) and $\ka=1/2$ (infrared analysis).  The scale $x=0.25$ roughly 
corresponds to the lowest lattice momenta.  See text for details.}
\end{figure}

There is however the possibility that the lattice results might be infrared 
constant.  In Fig.~\ref{fig:latconst}, as an illustrative example, we 
compare the results for the infrared divergent ghost propagator dressing 
function (numerical and infrared powerlaw, corresponding to $\ka=0.5$) with 
the numerical curve for $\G(0)=0.2$ and a fit form with $\ka=0.2$.  Also 
shown is the perturbative asymptotic result.  Importantly, for a wide range 
of values ($x\approx[0.25,10]$) the curves with $\G(0)=0.2$ and $\ka=0.2$ 
are indistinguishable.  Whilst this is only an illustration, it does give 
an alternative explanation to the lower lattice value of $\ka$ as 
corresponding to a nonzero value of $\G(0)$ whilst not contradicting the 
earlier infrared analysis.  Again, the deciding factor is that $x_{min}$ 
cannot be arbitrarily lowered.  Note also that since the lattice results 
with $x_{min}\sim0.25$ do see an infrared enhancement with $\ka=0.2$ rather 
than a constant behavior for the dressing function, this would presumably 
place an upper bound on the value of the boundary condition: roughly 
$\G(0)\leq0.2$.

To summarize, it would appear that in order to discriminate between the 
various infrared solutions, the lattice studies have not yet gone far 
enough into the infrared to give a clear signal that one has a particular 
value of the exponent $\ka$, or a particular boundary condition $\G(0)$.  
Such studies are planned \cite{mqgb} and using the analysis of the ghost 
\DS equation presented in this study may be of help in the extraction of 
the infrared behavior.
\section{\label{sec:five}Discussion of the results}
\setcounter{equation}{0}

In this Section, we shall present and discuss arguments for the possible 
interpretation of the numerical results.  To this purpose, we shall draw 
on various different ideas and concepts.  In order to aid clarity, we split 
the discussion into different parts.  The first part reiterates the 
identification of $\G(0)$ as a boundary condition.  The second then 
addresses the issue of the physical interpretation of such a boundary 
condition and its connection to the Gribov gauge fixing ambiguity.  The 
third part then continues by turning the arguments round, resulting in a 
conjecture that physical results may actually be independent of the value 
of the boundary conditions.  The final part then discusses an alternate 
interpretation of $\G(0)$ as being concerned with the renormalization 
scheme and why this is seemingly ruled out by the results here.

\subsubsection{$\G(0)$ as a boundary condition.}
That the various infrared solutions of the ghost \DS equation are 
delineated by some constant, $\G(0)$, should not come as a surprise.  
Indeed, this has been explicitly seen in the canonical approach to Coulomb 
gauge \cite{Epple:2007ut} and there is a similar situation in Landau gauge 
as discussed in the Introduction.  The \DS equations are derived as 
functional differential equations that can only \emph{relate} Green's 
functions and as such, require (functional) boundary conditions to finally 
specify a value for a given Green's function.  Clearly the solutions should 
agree with the asymptotic perturbative expansion in the vicinity of the 
renormalization point and the functions should be smooth and continuous for 
finite, non-zero momenta (a feature naturally present for an iterative 
numerical solution).  The remaining boundary condition in the case of the 
ghost \DS equation is then the value of $\G(0)$.  Normally, one would be 
considering a coupled system of \DS equations whereby the solution for the 
gluon propagator is also being sought and the issue of the existence of the 
boundary condition is invariably obscured by technical details (such as the 
truncation scheme); here we have utilized the general consensus about the 
Gribov form for the equaltime spatial gluon propagator to fix this sector 
so that the existence of the boundary condition as input to choosing the 
relevant solution is made explicit.  Indeed, this is particularly relevant 
in Coulomb gauge within the functional integral formalism, where the energy 
divergence of the ghost-loop and the necessary nonperturbative cancellation 
prohibits a direct back-coupling of the ghost to the gluon polarization.

\subsubsection{$\G(0)$ and the Gribov problem.}
Having established the requirement for a boundary condition, the important 
question is thus: is there a particular physical value?  To answer this 
question we obviously need to know to what physics the different solutions 
correspond to.  The most pertinent aspect of the numerical results 
presented here is that for a wide range of values ($0\leq\G(0)<0.3$), the 
UV solution and the ghost renormalization coefficient are unaltered and 
meaning that the boundary condition has nothing to do with either the 
perturbative region or the renormalization condition (at least 
conventionally where the renormalization scale is chosen to be 
perturbative).  We can thus conclude that the boundary condition, at least 
for the ghost \DS equation in isolation, is connected only to the 
nonperturbative (infrared) physics.  Additionally, that the various 
solutions exhibit a particular infrared behavior where as one goes from 
large to small momenta, the curves follow the dynamical solution until 
`forced' to deviate to the constant boundary value suggests that the 
boundary condition is connected to a freezing of the dynamics below a 
particular scale.

Let us now introduce some aspects of Coulomb gauge and the Gribov problem 
relevant to the discussion here.  In the functional approach to Coulomb 
gauge (within the first order formalism, but the same applies here), the 
system can be shown to reduce to the `would-be-physical' degrees of freedom 
\cite{Zwanziger:1998ez,Watson:2006yq,Reinhardt:2008pr}: that is, after 
resolving Gauss' law only two transverse spatial gluon degrees of freedom 
remain (in quantum electrodynamics these would be the physical photon 
polarization states) and the longitudinal, temporal, and ghost degrees 
of freedom cancel.  In the Gribov-Zwanziger confinement picture 
\cite{Gribov:1977wm,Zwanziger:1998ez,Zwanziger:1995cv}, the transverse 
spatial propagator (given by the Gribov formula here) is infrared 
suppressed and subsequently drops out of the physical spectrum.  The 
reduction to transverse spatial degrees of freedom has the Faddeev-Popov 
(FP) operator as a central 
element and the ghost propagator is precisely the expectation value of the 
inverse of this operator.  However, the known existence of zero modes of 
the FP operator and the associated Gribov ambiguities complicate matters 
\cite{Gribov:1977wm}.  The Gribov problem is a result of the fact that the 
gauge is not completely specified.  The lack of complete gauge fixing 
manifests itself in gauge-variant quantities (i.e., the propagators) and 
implies an ambiguity in the definition of the underlying functional 
integral.  
Given that the spatial gluon propagator in Coulomb gauge corresponds to 
the `would-be-physical' degrees of freedom (in the above sense), one 
might expect that the Gribov problem would 
be relatively unimportant -- we have, after all, the Gribov form as input; 
on the other hand, the ghost propagator would be 
entirely dependent on the resolution of the incomplete gauge-fixing 
(explicitly found to be the case in 1+1 dimensions 
\cite{Reinhardt:2008ij}).  This scenario is the case here with a fixed 
spatial gluon propagator (the Gribov form) and a ghost that must be 
specified with further input but that does not directly back-couple to the 
gluon.  

The first part of the argument for a connection between the Gribov problem 
and the boundary condition $\G(0)$ goes as follows.  In the presence of 
the Gribov 
ambiguity, the form of the \DS equations (at least in more than one spatial 
dimension for Coulomb gauge) is unchanged \cite{Zwanziger:2003cf}.  The 
implication is that the dynamical behavior of the solutions would not be 
affected by the existence of Gribov copies and this is consistent 
with the characteristic pattern of the results here 
(Fig.~\ref{fig:ghres0}) where with various values of 
$\G(0)$ (at least for the range $0\leq\G(0)<0.3$) as $\G(0)$ is 
lowered, the solutions follow the same dynamical curve further into the 
infrared before freezing out to the value $\G(0)$ imposed as the boundary 
condition.

The second part of the argument connecting $\G(0)$ and the Gribov ambiguity 
concerns the following observation.  
Generally, when (improperly) integrating over the gauge group, 
gauge-dependent functional integrals (propagators) would be suppressed by 
averaging over the gauge copies.  
This is an important point: gauge copies either have the same 
phase in group space when contributing to the functional integral in which 
case the integration over the gauge copies is merely an unobservable 
constant in the normalization; or if the gauge copies have a different 
phase then the integration over the group space can only reduce a 
gauge-dependent functional integral.  Gribov copies are accompanied by zero 
modes of the Faddeev-Popov operator, which for Coulomb gauge reads
\be
-\s{\div}{\vec{D}^{ab}[\vec{A}]},\;\;\;\;
\vec{D}^{ab}[\vec{A}]=\de^{ab}\div-gf^{acb}\vec{A}^c(x).
\ee
Perturbatively, in the sense of an expansion in the gauge field ($\vec{A}$) 
around zero there are no Gribov copies such that the perturbative regime 
would be insensitive.  The ghost propagator results are insensitive 
to $\G(0)$ at large momenta where perturbation theory is valid 
(as explicitly demonstrated previously).  In the nonperturbative case, for 
low momenta, there are `more' Gribov copies (we do not know how to count 
the number, but from the form of the Faddeev-Popov operator, the large 
fields $\vec{A}$ within the covariant derivative that account for the 
existence of Gribov copies are large in the sense of their relationship to 
the spatial differential operator, which in momentum space translates, 
loosely speaking, to the spatial momentum).  There are two plausible 
signals for the possible presence of Gribov copies within the functional 
integral for the ghost propagator arising from improper gauge averaging: 
1) below a certain momentum scale, the 
averaging over the gauge group induced by the existence of Gribov copies 
might serve to freeze the functional integral to a specific value and 
2) 
for successively lower momenta, the difference between a functional integral 
affected by Gribov copies and one that isn't would increase.  
The results in this study show that for finite $\G(0)$, there is precisely 
such a freezing below a particular scale whereas for $\G(0)=0$ there isn't.  
Also, the difference between the various solutions with finite and zero 
values of $\G(0)$ does increase with successively lower momenta.  Now, 
$\G(0)$ is a boundary condition specifying a particular allowed solution of 
the \DS equation and thereby implicitly selecting a definition of the ghost 
propagator functional integral from the various possibilities given that 
there is an ambiguity in the gauge fixing.  That for different values of 
$\G(0)$, the results explicitly exhibit the characteristics that one would 
expect in the possible presence of Gribov copies suggests that $\G(0)$ is 
indeed intimately connected to the Gribov problem.  The case $\G(0)=0$ 
would appear to correspond to that implicit definition of the functional 
integral for the ghost where the Gribov copies have no effect, i.e., the 
curve is purely dynamical and there is no freezing in the infrared.  For 
finite $\G(0)$, the dynamical curve freezes below a certain scale indicating 
the influence of Gribov copies and their averaging.

There are two further arguments connecting $\G(0)$ and the Gribov problem 
that are known in Coulomb gauge: Gribov's original work \cite{Gribov:1977wm} 
and the case of $1+1$-dimensions \cite{Reinhardt:2008ij}.  In 
Ref.~\cite{Gribov:1977wm}, the perturbative aspects of the ghost propagator 
were discussed.  In the language used here, in order to ensure the absence 
of gauge copies (in the sense that the FP operator be positive definite), 
the restriction $\G(0)\geq0$ was introduced.  In the case of 
$1+1$-dimensional Coulomb gauge (where 
because the `physical' transverse spatial propagator cannot exist, the 
system reduces to purely a discussion of the gauge-fixing and Gribov 
copies) it was seen explicitly that the ghost propagator functional integral 
after resolving the Gribov problem was that with effectively the lowest 
allowed positive value of the boundary condition (due to the spatial 
compactification of the underlying manifold employed there, one could not 
assign a value of $\G(x)$ for $x=0$).  Functional integrals constructed to 
include Gribov copies had different boundary conditions, but obeyed the same 
\DS equation.

One objection to the above assertion that $\G(0)=0$ corresponds to the 
situation where Gribov copies have no effect is of course that the 
lattice data may be consistent with the solution for $\G(0)\sim0.2$ as 
discussed earlier.  That the ghost dressing function is infrared finite 
seems to be the case in Landau gauge on the lattice (see for example 
Ref.~\cite{Cucchieri:2010xr} and references therein).  However, it could 
be that the different results are not in conflict but are results for 
different problems.  Gribov copies are the consequence of an incomplete 
gauge fixing, which manifests itself in the eigenspectrum 
of the Faddeev-Popov operator and this eigenspectrum is obviously dependent 
on the boundary conditions one assigns to the eigenfunctions.  Namely, the 
Gribov problem may be different for a compact manifold (the lattice with 
periodic boundary conditions) or in the continuum (with a flat 
Minkowski/Euclidean metric and boundary conditions at infinity) and also 
different when comparing Coulomb and Landau gauges 
\cite{Greensite:2010hn}.  When inserted into the functional integral, it is 
then currently an open question as to what effect this might have.

\subsubsection{$\G(0)$ as an implicit gauge choice.}
Another way to view the connection between $\G(0)$ and the Gribov problem 
is that it can be conjectured that the additional (nonlocal) gauge 
fixing required to eliminate Gribov copies is linked to the discussion of 
a specific choice for $\G(0)$.  However, turning the argument around, one 
could interpret $\G(0)$ as a choice of gauge, as was proposed in 
Ref.~\cite{Maas:2009se}.  More properly, the interpretation of $\G(0)$ 
would be as a choice of gauge completion: the overall gauge choice 
(Landau or Coulomb gauge, for example) is required to have a well-defined 
perturbative propagator and because the dynamical solution to the \DS 
equation must be smoothly and continuously connected to this perturbative 
expression in the ultraviolet, then the lack of complete gauge fixing does 
not mean that the Green's function is ill-defined, rather that there exist 
multiple solutions and the choice of gauge completion defines the 
nonperturbative component of the propagator in the sense that a specific 
solution (corresponding to a particular value of the boundary condition) is 
chosen.  Indeed, the interpretation of the boundary condition $\G(0)$ as an 
implicit choice of gauge is rather convenient since as long as one can show 
that the different choices are physically equivalent, one can choose the 
gauge so as to make specific calculations more straightforward.  That is to 
say that whilst the elimination of Gribov copies may indeed provide a 
definite value for $\G(0)$, the arbitrariness is to do with the gauge 
fixing and one may specify any gauge one desires.  This provides a possible 
answer to the original question about the existence of a particular 
physical value for the boundary condition $\G(0)$ -- there may be no unique 
\emph{physical} value, just as propagators are different in different 
gauges.  Furthermore, issues such as the confinement mechanism may appear 
different for different choices of $\G(0)$.  The task would of course be to 
eventually show that the same physical results can be obtained with 
different boundary conditions.  We shall briefly discuss the case $\G(0)=0$ 
below.

\subsubsection{$\G(0)$ as a renormalization scheme.}
As a final remark here, another interpretation of the boundary condition is 
that the value of $\G(0)$ is a choice of the renormalization scheme to 
absorb constants into the renormalization coefficient $Z_c$, and which 
again is a free choice (and which has no impact on the perturbative 
renormalization).  The Gribov-Zwanziger confinement scenario 
\cite{Gribov:1977wm,Zwanziger:1998ez,Zwanziger:1995cv} motivates the choice 
$\G(0)=0$.  However, that the renormalization coefficient $Z_c$ is seen to 
be stable for a range of values, $\G(0)$ (see Fig.~\ref{fig:zc}), would 
seem to rule out this 
interpretation of $Z_c$ for this aspect of the Gribov-Zwanziger scenario in 
Coulomb gauge.
\section{\label{sec:six}Critical ghost confinement and infrared slavery}
\setcounter{equation}{0}
As has been seen, when the boundary condition $\G(0)=0$, the ghost 
propagator dressing function behaves like $1/|\vec{k}|$ in the infrared.  
This is of course the result for the so-called critical solutions obtained 
in the canonical formalism and provides a very simple picture for many 
aspects of confinement \cite{Reinhardt:2008ek,Reinhardt:2009ks}.  In the 
functional formalism, the infrared critical (powerlaw) ghost solution 
provides an important possible addition to this list, as we shall now 
motivate.

The ghost and spatial gluon propagators have been discussed so far, but in 
the functional approach to Coulomb gauge there is a third propagator -- the 
temporal gluon propagator corresponding to the temporal component of the 
gluon field ($\si\equiv A^0$).  In the canonical approach to Coulomb gauge, 
one starts with Weyl gauge ($\si=0$) and then imposes Coulomb gauge 
\cite{Christ:1980ku,Feuchter:2004mk}, so the temporal gluon propagator does 
not exist in this sense (the analogous quantity to the temporal gluon 
propagator in the canonical formalism is the non-Abelian color potential).  
Within the functional formalism in the heavy quark limit, it has been shown 
that when the temporal gluon propagator diverges like $1/\vec{k}^4$ 
(or equivalently, that the dressing function goes like $1/\vec{k}^2$) in 
the infrared, one directly obtains a linear rising potential 
(and bound states) for color singlet quark-antiquark and diquark 
(for $N_c=2$ colors) configurations with no finite energy colored states 
permitted and corresponding precisely to the old infrared slavery picture 
of confinement \cite{Popovici:2010mb}.  The lattice results for the 
temporal gluon propagator do exhibit this infrared behavior 
\cite{Quandt:2008zj,Cucchieri:2000gu}.

The decomposition of the temporal gluon propagator and proper two-point 
function reads \cite{Watson:2007vc}:
\be
W_{\si\si}^{ab}(k)=
\imath\de^{ab}\frac{D_{\si\si}(k_0^2,\vec{k}^2)}{\vec{k}^2},\;\;\;\;
\G_{\si\si}^{ab}(k)=
-\imath\de^{ab}\vec{k}^2\G_{\si\si}(k_0^2,\vec{k}^2),\;\;\;\;
D_{\si\si}(k_0^2,\vec{k}^2)\G_{\si\si}(k_0^2,\vec{k}^2)=1.
\ee
In addition, the decomposition of the proper spatial gluon two-point 
function (the spatial gluon polarization) is
\be
\G_{AAij}^{ab}(k)=-\imath\de^{ab}
\left[k^2t_{ij}(\vec{k})\G_{AA}(k_0^2,\vec{k}^2)
+k_0^2\frac{k_ik_j}{\vec{k}^2}\ov{\G}_{AA}(k_0^2,\vec{k}^2)\right],
\ee
where the dressing function $\G_{AA}$ is the inverse of $D_{AA}$ and 
$\ov{\G}_{AA}$ is the longitudinal dressing function component of the 
polarization (in Coulomb gauge, this quantity is emphatically not 
vanishing unlike in Landau gauge).  Importantly, the \ST identity for the 
two-point functions tells us that \cite{Watson:2008fb}
\be
\G_{\si\si}(k_0^2,\vec{k}^2)=
\ov{\G}_{AA}(k_0^2,\vec{k}^2)\left[\G_{\ov{c}c}(\vec{k}^2)\right]^2
\ee
(we replace the subscripts for $\G_{\ov{c}c}$ to avoid confusion).  On 
general grounds, it can be shown that $D_{\si\si}$ (and consequently, 
$\G_{\si\si}$) must have a nontrivial instantaneous (energy independent, 
or $\sim\de(x_1^0-x_2^0)$ in configuration space) component 
\cite{Cucchieri:2000hv}.  In fact, lattice results indicate that there is 
only a very weak energy dependence for $D_{\si\si}$ \cite{Quandt:2008zj} 
and so it appears that neglecting the non-instantaneous component is a 
good approximation.  In other words
\be
D_{\si\si}(k_0^2,\vec{k}^2)=D_{\si\si}(\vec{k}^2),\;\;\;\;
\G_{\si\si}(k_0^2,\vec{k}^2)=\G_{\si\si}(\vec{k}^2),\;\;\;\;
\ov{\G}_{AA}(k_0^2,\vec{k}^2)=\ov{\G}_{AA}(\vec{k}^2).
\ee
Assuming multiplicative renormalizability and assigning the renormalization 
coefficient $\sqrt{Z_\si}$ to the temporal gluon field $\si$ as before, 
since both the longitudinal and transverse components of the spatial 
polarization share the same renormalization coefficient, then we 
immediately see that
\be
Z_\si=Z_AZ_c^2
\ee
as a consequence of the \ST identity, as has been known for some time 
\cite{Zwanziger:1998ez}.  With \eq{eq:rginv1}, the product $g^2D_{\si\si}$ 
is thus a renormalization group invariant, just as 
$g^2D_{AA}D_{\ov{c}c}^2$ is.  Taking the infrared powerlaw form for the 
ghost propagator dressing function (noting the scaling of the variables 
with $m^2$), one infers that
\be
g^2D_{\si\si}(\vec{k}^2)\stackrel{\vec{k}^2\rightarrow0}{\sim}
\frac{m^2}{\vec{k}^2\ov{\G}_{AA}(\vec{k}^2)}
\ee
meaning that if $\ov{\G}_{AA}$ is a constant in the infrared, one has 
exactly the temporal gluon interaction required for a linear rising 
potential in Coulomb gauge.  Recall that the lattice results for the 
spatial gluon propagator indicate that the Gribov scale $m$ is not 
dependent on the renormalization scale \cite{Burgio:2008jr} and one had 
the approximate relationship $m\sim2\sqrt{\si_w}$ where $\si_w$ is the 
Wilsonian string tension.  The numerator factor $m^2$ arising from the 
Gribov formula would be thus entirely consistent with the coefficient of 
the linear rising potential discussed in Ref.~\cite{Popovici:2010mb}.  
Also recall that in Coulomb gauge, the spatial gluon propagator corresponds 
to the `would-be-physical' transverse degrees of freedom which indeed 
suggests that the 
Gribov scale should be related to a physically observable scale.  So, if 
$\ov{\G}_{AA}$ were to be a constant in the infrared, there would be a 
direct connection between the spatial gluon propagator and the confinement 
potential even though the spatial gluon propagator is infrared vanishing 
(as suggested by the Gribov-Zwanziger confinement scenario 
\cite{Gribov:1977wm,Zwanziger:1998ez,Zwanziger:1995cv}).

That $\ov{\G}_{AA}$ is a constant can be argued by appealing to the 
dimensional type of infrared analysis employed earlier for the ghost \DS 
equation and general arguments for \DS results seen in Landau gauge.  In 
terms of renormalized quantities and recognizing that the renormalization 
coefficients are dependent on the regularization scale, $\La$, the ghost 
\DS equation reads schematically
\be
\G_{\ov{c}c}^R(\vec{k}^2;\mu,\ov{g})=
Z_c(\mu,\ov{g};\La)
+\tilde{Z}_A(\mu,\ov{g};\La)\Si_{\ov{c}c}(\vec{k}^2;\mu,\ov{g};\La).
\ee
Now, with the result that $\tilde{Z}_A=1$, the self-energy integral must 
separate into two components, i.e.,
\be
\Si_{\ov{c}c}(\vec{k}^2;\mu,\ov{g};\La)\rightarrow
\Si_{\ov{c}c}^{fin}(\vec{k}^2;\mu,\ov{g})
+\Si_{\ov{c}c}^{div}(\mu,\ov{g};\La)
\ee
such that the equation can be renormalizable.  The earlier dimensional 
analysis for the infrared could then be performed without reference to the 
(large) scale $\La$ and $\Si_{\ov{c}c}^{div}$ after a straightforward 
subtraction of the equation for $\vec{k}^2=\mu$.  Then, counting the powers 
of $\vec{k}^2$ led directly to the powerlaw exponent if one assumed the 
absence of the infrared constant solution and disregarded the Gribov scale 
$m$.  In the case of the \DS equation for $\ov{\G}_{AA}$, one has the 
similar schematic form
\be
\ov{\G}_{AA}^R(\vec{k}^2;\mu,\ov{g})=
Z_A(\mu,\ov{g};\La)
+\sum_{i}Z_i(\mu,\ov{g};\La)\Si_i(\vec{k}^2;\mu,\ov{g};\La)
\ee
but with many more loop self-energy terms (labeled with the index $i$).  
The renormalization coefficients $Z_i$ are the coefficients for the dressed 
vertices that enter the corresponding loop integrals and can be explicitly 
derived from the \ST identities \cite{Watson:2008fb}.  The important 
information is that only the spatial ghost-gluon vertex coefficient is 
trivial ($\tilde{Z}_A=1$).  Thus, only the ghost loop of the \DS equation 
for $\ov{\G}_{AA}$ has a form similar to the ghost self-energy above, as is 
also the case in Landau gauge.  However, in Coulomb gauge this loop is 
explicitly energy divergent \emph{and must cancel}.  Therefore, one cannot 
separate the finite ($\La$-independent) and divergent 
($\vec{k}^2$-independent) parts of any of the loop integrals $\Si_i$ and 
the dimensional analysis, which relies on this decomposition, cannot hold.  
This argument would also be true for the \DS equations for $\G_{AA}$ 
(the transverse part of the polarization).  Actually, the lack of such a 
separation between infrared and ultraviolet parts of the self-energy terms 
is another of the difficult technical problems associated with the 
functional approach to Coulomb gauge and is an explicit demonstration of 
the difference to Landau gauge (where it is the ghost loop that is the 
important infrared contribution to the polarization 
\cite{von Smekal:1997vx}).

In the absence of the dimensional analysis, the infrared character of the 
\DS equations cannot be directly inferred.  However, one can use the 
experience gained from Landau gauge to anticipate the likely outcome.  It 
has been known for a long time that in Landau gauge, there exist potential 
quadratic UV divergences in the tensor component of the polarization that 
is proportional to the metric, which here would be the transverse part of 
the spatial gluon polarization (i.e., the \DS equation for $\G_{AA}$) 
\cite{Brown:1988bn} (see also Ref.~\cite{Fischer:2008uz} and references 
therein).  Thus, one would naively expect terms in the unrenormalized 
equation for the dimensionless dressing function that behave like 
$\La/\vec{k}^2$, but which should cancel with a proper regularization 
procedure (and ignoring in this case the energy dependence).  After 
regularization and renormalization 
(which doesn't alter the $\vec{k}^2$-dependence), such terms can only be of 
the form $m^2/\vec{k}^2$ or $\mu/\vec{k}^2$, with any other terms being 
infrared subleading.  Thus, one would expect an infrared behavior for 
$D_{AA}=\G_{AA}^{-1}$ of the form
\be
D_{AA}\stackrel{\vec{k}^2\rightarrow0}{\sim}\vec{k}^2,
\ee
which is exactly the case for the Gribov formula (and observed on the 
lattice).  For the longitudinal part of the polarization, one would expect 
logarithmic UV divergences and in the presence of the additional Gribov 
scale $m$, this would indeed favor a constant type of infrared solution, 
just as is the case for the dressing functions in the fermion sector of QCD 
(see for example Ref.~\cite{Alkofer:2002bp}):
\be
\ov{\G}_{AA}\stackrel{\vec{k}^2\rightarrow0}{\sim}\mbox{const}.
\ee
In other words, one can justify that $\ov{\G}_{AA}$ is indeed constant in 
the infrared such that with the ghost boundary condition $\G(0)=0$, the 
temporal propagator would be precisely that required for a linearly rising 
confinement potential.

Obviously, the discussion of this section is somewhat speculative and is 
certainly intended more to discuss the plausibility of a simple confinement 
scenario in Coulomb gauge than as a quantitative calculation.  
One example of a technical complication that might arise in such a 
quantitative study is that the \ST identities are in general deformed by 
the presence of a UV cutoff scale $\La$: in Landau gauge, this has been 
explicitly studied in Ref.~\cite{Fischer:2008uz} (and references therein).  
However, 
what should be amply clear is that the confinement mechanism in Coulomb 
gauge is certainly dependent on the ghost boundary condition $\G(0)$, 
underscoring the necessity of pursuing the  understanding of the ghost 
propagator in the infrared further.
\section{\label{sec:seven}Summary and outlook}
\setcounter{equation}{0}

In this study, the ghost \DS equation in Coulomb gauge has been considered 
numerically using lattice input for the spatial gluon propagator.  It is 
demonstrated that the dynamical solution must be explicitly supplemented by 
a nonperturbative boundary condition (the value of the inverse ghost 
propagator dressing function at zero spatial momentum).  With various 
values of this boundary condition, the solutions have a characteristic 
behavior: in the ultraviolet, all solutions lie on top of each other and 
agree with the asymptotic leading order perturbative result; going from 
high to low momenta, the solutions follow the common dynamical curve into 
the infrared until freezing out at a point determined by the boundary 
condition.  In this way, it is seen that both critical (powerlaw) and 
subcritical (infrared finite) solutions exist.  It was seen that the 
critical solution is characterized in the infrared by the exponent 
$\al=1/2$.  The ghost renormalization coefficient was observed to be 
largely independent of the boundary condition.  A qualitative comparison to 
lattice results for the ghost propagator dressing function was 
illustrated, with the conclusion that it would be desirable to have 
lattices that go further into the infrared in order to discriminate between 
the critical and subcritical solutions.  The characteristic pattern of the 
solutions for different boundary conditions can be naturally interpreted in 
terms of the Gribov gauge fixing ambiguity and it is conjectured that if 
the boundary condition is related to a choice of gauge fixing completion, 
then physical (gauge invariant) results should not be dependent on the 
choice.  Plausibility arguments are put forward that connect the critical 
solution for the ghost propagator dressing function with the temporal gluon 
propagator and the confinement mechanism.

Whilst the aim of this study was to demonstrate the explicit requirement 
for a nonperturbative boundary condition to supplement the dynamical 
solution of the ghost \DS equation in Coulomb gauge and to present 
numerical results, the characteristic behavior that was found led to two 
primary conjectures.  Clearly then, the task will be to investigate 
further.  The first conjecture is that the boundary condition studied here 
is connected to the resolution of the Gribov problem in Coulomb gauge.  
Work is currently underway to study this connection explicitly 
\cite{prog}.  The second conjecture concerns the connection of the ghost 
propagator to confinement via the temporal and longitudinal spatial gluon 
proper two-point functions courtesy of the \ST identities.  This is a 
formidable task because of the inherent technical difficulties of solving 
the \DS equations within the noncovariant Coulomb gauge functional 
formalism; however, these technical problems are being steadily overcome 
and one may hope for concrete results within a reasonable timeframe.

\begin{acknowledgments}
The authors gratefully acknowledge useful discussions with G.~Burgio and 
M.~Quandt about the lattice results and D.~Campagnari for suggestions 
leading to the fit form \eq{eq:gfit}.  This work has been supported by the 
Deutsche Forschungsgemeinschaft (DFG) under contracts no. DFG-Re856/6-2,3.
\end{acknowledgments}

\end{document}